%% file: main.tex
\documentclass[sigconf]{acmart}
\usepackage[utf8]{inputenc}
\usepackage[T1]{fontenc}
\usepackage[normalem]{ulem}
\usepackage{amsmath}
\usepackage{bbm}
\usepackage{xcolor}
\usepackage{natbib}
\usepackage{graphicx}
\usepackage[skip=-1pt]{caption}
\usepackage{subcaption}
\usepackage{amsmath}
\usepackage{makecell}
\usepackage{amsfonts}
\usepackage{hyperref}
\usepackage[inline]{enumitem}
\usepackage{booktabs}
\usepackage{multirow}
\usepackage{mathtools}
\usepackage{multicol}
\usepackage{amsmath, bm}
\usepackage{stmaryrd}
\usepackage{longtable}
\usepackage{autobreak}
\usepackage{breqn}
\usepackage{titlesec}
\usepackage{scalerel}
\usepackage[ruled,vlined,english,onelanguage]{algorithm2e}
\usepackage{cleveref}

\theoremstyle{definition}

\newcommand{\ie}{\emph{i.e.}}
\newcommand{\eg}{\emph{e.g.}}

\newcommand{\vs}{\emph{vs. }}

\graphicspath{ {./fig/}  }  

\copyrightyear{2022} 
\acmYear{2022} 
\setcopyright{acmcopyright}\acmConference[SIGIR '22]{Proceedings of the 45th International ACM SIGIR Conference on Research and Development in Information Retrieval}{July 11--15, 2022}{Madrid, Spain}
\acmBooktitle{Proceedings of the 45th International ACM SIGIR Conference on Research and Development in Information Retrieval (SIGIR '22), July 11--15, 2022, Madrid, Spain}
\acmPrice{15.00}
\acmDOI{10.1145/3477495.3532007}
\acmISBN{978-1-4503-8732-3/22/07}

\begin{document}
\fancyhead{}

\title{Joint Multisided Exposure Fairness for Recommendation}

\author{Haolun Wu}
\authornote{Both authors contributed equally to this research.}
\authornote{Corresponding author.}
\affiliation{
  \institution{McGill University}
  \city{Montréal}
  \country{Canada}
}
\email{haolun.wu@mail.mcgill.ca}

\author{Bhaskar Mitra}
\authornotemark[1]
\affiliation{
  \institution{Microsoft}
  \city{Montréal}
  \country{Canada}
}
\email{bmitra@microsoft.com}

\author{Chen Ma}
\authornotemark[2]
\affiliation{
  \institution{City University of Hong Kong}
  \country{Hong Kong SAR}
}
\email{chenma@cityu.edu.hk}

\author{Fernando Diaz}
\affiliation{
  \institution{Canadian CIFAR AI Chair \& Google}
  \city{Montréal}
  \country{Canada}
}
\email{diazf@acm.org}

\author{Xue Liu}
\affiliation{
  \institution{McGill University}
  \city{Montréal}
  \country{Canada}
}
\email{xueliu@cs.mcgill.ca}

\input{abstract.tex}

\ccsdesc[500]{Information systems~Recommender systems}

\keywords{Fairness-aware Recommendation; Multisided Fairness; Exposure Fairness; Recommender System}

\maketitle

\input{intro.tex}
\input{related.tex}
\input{prelims.tex}

\input{jme.tex}
\input{analysis.tex}

\input{optimization.tex}
\input{conclusion.tex}

\bibliographystyle{ACM-Reference-Format.bst}
\bibliography{references.bib}
\end{document}

%% file: abstract.tex
\begin{abstract}
Prior research on exposure fairness in the context of recommender systems has focused mostly on disparities in the exposure of individual or groups of items to individual users of the system. The problem of how individual or groups of items may be systemically under or over exposed to groups of users, or even all users, has received relatively less attention. However, such systemic disparities in information exposure can result in observable social harms, such as withholding economic opportunities from historically marginalized groups (\emph{allocative harm}) or amplifying gendered and racialized stereotypes (\emph{representational harm}). Previously, \citet{diaz2020evaluating} developed the \emph{expected exposure} metric---that incorporates existing user browsing models that have previously been developed for information retrieval---to study fairness of content exposure to individual users. We extend their proposed framework to formalize a family of exposure fairness metrics that model the problem jointly from the perspective of both the consumers and producers. Specifically, we consider group attributes for both types of stakeholders to identify and mitigate fairness concerns that go beyond individual users and items towards more systemic biases in recommendation. Furthermore, we study and discuss the relationships between the different exposure fairness dimensions proposed in this paper, as well as demonstrate how stochastic ranking policies can be optimized towards said fairness goals.

\end{abstract}

%% file: intro.tex
\section{Introduction}
\label{sec:intro}

Online information access systems, like recommender systems and search, mediate what information gets exposure and thereby influence their consumption at scale.
There is a growing body of evidence~\citep{kay2015unequal, singh2018fairness, diaz2020evaluating} that information retrieval (IR) algorithms that narrowly focus on maximizing ranking utility of retrieved items may disparately expose items of similar relevance from the collection.
Such disparities in exposure outcome raise concerns of algorithmic fairness and bias of moral import~\citep{friedman1996bias}, and may contribute to both \emph{representational} and \emph{allocative} harms~\citep{crawford2017trouble}.
Representational harms may manifest by reinforcing negative stereotypes---\eg, disproportionately suggesting arrest record searches for black-identifying first names in online ad delivery systems~\citep{sweeney2013discrimination}---and perpetuating inequities in representation of women and other historically marginalized peoples in different occupational roles~\citep{kay2015unequal}.
Similarly, allocative harms may occur when disparate exposure in retrieved results lead to unfair allocation of economic opportunities~\citep{singh2018fairness}---\eg, in IR systems that aim to match employers with potential candidate employees, and for content-producers on the web who depend on search traffic for ad-based monetization.

The \emph{probability ranking principle}~\citep{robertson1977probability}, which suggests that retrieved items should be ordered in decreasing probability of relevance, has been a guiding principle in many IR model development.
In contrast, by viewing IR systems as mediators of exposure, \citet{diaz2020evaluating} propose the \emph{principle of equal expected exposure} that requires IR systems to provide equal exposure in expectation to items of comparable relevance for a given information need.
Furthermore, they define the expected exposure metric as a deviation between actual system exposure and ideal target exposure for individual (or groups of) items to individual users.
While \citet{diaz2020evaluating} consider group attributes only on the producer-side, \citet{burke2017multisided} and \citet{ekstrand2021fairness} argue that it is important to consider group attributes on both the consumer-side and producer-side, and that the joint consideration of group attributes on both sides can reveal new fairness concerns.
However, these previous works do not provide a formal formulation of joint multisided exposure (JME) fairness, which is exactly where our current work fits in.
We extend the expected exposure metric proposed by \citet{diaz2020evaluating} to incorporate group attributes on both the consumer and producer side and formally define a broader set of fairness concerns that we believe should receive consideration for developing fair recommendation algorithms.

These distinct fairness concerns may be best explained through specific examples.
Consider a job portal that recommends potential employment opportunities to job seekers.
In this recommendation scenario, the employers represent the producer-side and the job seekers consume the recommendations produced by the system.
A responsible recommendation algorithm in this context must grapple with several fairness concerns.
Firstly, any individual user of the system must have fair exposure to all relevant jobs that they may qualify for.
We refer to this as individual-user to individual-item fairness, or II-F to be more concise.
If we group similar job roles by salary into, say, high and low paying job groups, then the system must also ensure that an individual job candidate does not disproportionately get recommended either low or high paying jobs.
We classify such fairness concerns as individual-user to group-of-items fairness, or IG-F in short.
If we now group the job seekers---for example, based on race or gender---then the system must also guarantee that qualified candidates get fair exposure to all relevant individual job opportunities independent of their demographic attributes.
We refer to this as group-of-users to individual-item fairness, or GI-F.
Extending the framework further, we consider the scenario where job seekers are grouped by demographic attributes, like race and gender, and employment opportunities by salary.
This raises a group-of-users to group-of-items fairness (GG-F) consideration such as ensuring that women or racially marginalized groups are not disproportionately recommended low-paying jobs.
Another interesting example of the GG-F concern involves grouping both job seekers and employers by demographic, such as race, where an unfair recommender system may, for example, disproportionately recommend Black employees to Black-owned businesses and similarly match White employees with White employers, and codify \emph{algorithmic segregation} much like what \citet{benjamin2019race} calls the ``New Jim Code''.
Finally, the recommender system should also ensure that any individual jobs or groups of jobs (\eg, grouped by race or gender attributes of business owners) are not systemically under or over exposed to all users, which we refer to as all-users to individual-item fairness (AI-F) and all-users to group-of-items fairness (AG-F), respectively.
For example, an unfair recommender system may systemically under-expose jobs at businesses owned by historically marginalized groups which according to our definition would fall under AG-F concerns.

While II-F and IG-F has been previously studied by \citet{diaz2020evaluating}, to the best of our knowledge, this is the first work to formally define and study a family of exposure fairness considerations jointly from the perspective of both the consumers and producers.
In summary, the key contributions of this work are as follows:\vspace{-1mm}
\begin{enumerate}
    \item We formalize and compare a family of JME-fairness measures that deal with different types of systemic biases in content exposure from recommender systems.
    \item We demonstrate how each of the six JME-fairness metrics can be decomposed into their corresponding disparity and relevance components, and study their trade-offs.
    \item We demonstrate how stochastic ranking policies can be optimized towards specific JME-fairness goals.
\end{enumerate}

%% file: related.tex
\section{Related work}
\label{sec:related}

\subsection{Fairness in recommendation}
\label{sec:related-fairness}
Fairness in recommendation has received increasing attention in the IR community recently.
Although there is no unified notion of fairness, one of the criteria is to model from the perspective of different stakeholders~\citep{ekstrand2021fairness}, which may include among others:
\begin{enumerate*}[label=(\roman*)]
    \item consumers (\ie, users who seek content),
    \item producers (\ie, users who create or publish content),
    \item information subjects (when retrieved items correspond to individuals---\eg, recommending candidates for jobs), and
    \item other side stakeholders who may be impacted even if they do not directly interact with the retrieval system (\eg, historically marginalized groups who may suffer from representational harms by information access platforms).
\end{enumerate*}

Even for a single stakeholder, different fairness dimensions are important to consider that if unchecked may result in different harmful outcomes.
For example, for consumers it is important to consider fairness in quality of service, such as ensuring that consumers belonging to different demographic groups experience comparable retrieval quality.
\citet{ekstrand2018all} and \citet{neophytou2022revisiting} demonstrate that recommender performance can vary across demographic groups and \citet{mehrotra2017auditing} report similar observations in the context of web search, another important retrieval scenario.
On similar lines, \citet{wu2021multifr} explicitly model the consumer-sided fairness in quality of service as the difference of Normalized Discounted Cumulative Gain (NDCG) between group of users with different demographic attributes which they then try to minimize during the model optimization process.

Apart from quality of service, fairness concerns may also arise with respect to what content consumers are exposed to~\citep{singh2018fairness,diaz2020evaluating}.
For example, when retrieved results correspond to economic opportunities, unfair distribution of exposure can result in allocative harm.
Take the case of a job recommender system, in addition to ensuring that different demographic groups of users receive results of comparable relevance, it is also critical that there are no systemic disparities in exposure to high and low paying jobs across demographics~\citep{burke2017multisided}.
Previous works have modeled fairness from the perspective of fair allocation and utilize some notions in social welfare, such as Envy-freeness~\cite{PatroBGGC20FairRec} and Least Misery~\cite{LinZZGLM17FairGroup}.
Disparate exposure can similarly raise concerns of stereotyping of consumers, say in case of a news recommender system whose results may reflect historical gender-based stereotyping~\citep{wu2021fairrec}.
Several previous works~\citep{zhang2018mitigating, BoseH19compFair, rekabsaz2021societal} have explored adversarial approaches, based on domain-confusion~\citep{tzeng2014deep, cohen2018cross}, to learn representations that conceal information about protected attributes, such as race or gender.

Other works focus on the fairness on the producer side, which is represented by the systematic exposure disparities~\cite{asia:equity-of-attention,singh:fair-pg-rank,diaz2020evaluating, ge2022explainable} across content providers.
This line of research mainly aims to ensure the equity of exposure for individual producers or group of producers with sensitive attributes.
For instance, \citet{ZehlikeB0HMB17FAIR} model producer-sided fairness in a top-$k$ ranking problem and guarantee a minimum proportion of items from the protected group in every prefix of the top-$k$ ranking. \citet{asia:equity-of-attention} aim to achieve amortized fairness of attention by making exposure proportional to relevance through integer linear programming.
\citet{singh2018fairness} later propose a more general framework which can achieve individual fairness and group fairness on the producer side concurrently.
In the learning-to-rank~\citep{liu2011learning} setting, \citet{singh:fair-pg-rank} propose a policy learning approach for optimizing ranking models while satisfying fairness
of exposure constraints, while \citet{diaz2020evaluating} develop a direct supervised approach using Gumbel sampling.
Recently, \citet{oosterhuis2021computationally} proposes computationally efficient optimization of Plackett-Luce ranking models for fairness.

It is important to ensure fairness for multiple stakeholders in online platforms for avoiding the super-star economy~\cite{MehrotraMBL018fairmarketplace, baranchuk2011economics} and building a healthy marketplace, as well as to ensure that improving fairness for one set of stakeholders do not negatively impact the utility of the other sides.
Several previous works~\citep{burke2018balanced, MehrotraMBL018fairmarketplace, suhr2019ridehailing, wu2021multifr} have explored multisided fairness definitions within recommendation tasks.
However, among these related works, with the exception of \citet{burke2018balanced}, the remaining focus on the fairness of quality of service, and not of exposure, on the consumer side.
The specific problem of two-sided exposure fairness has received limited attention in the community, despite the fact that it has been previously suggested by \citet{burke2017multisided} and \citet{ekstrand2021fairness}. Our work thus aims to fill that gap by presenting a formal formulation that extends previous works on user browsing model based exposure fairness~\citep{diaz2020evaluating}.

As the literature on the fairness of recommendation has recently grown significantly, we point our readers to \citep{ekstrand2021fairness} for a broader overview of fairness concerns and mitigation approaches in information access systems.

\subsection{Ranking with stochastic policy}
\label{sec:related-stochastic}
Most of the early works in learning-to-rank for IR~\citep{liu2011learning} focus on deterministic ranking policies that produce static ordering of items given a user (in case of recommendation) or a query (in case of search).
Motivated by \citet{PandeyROCC05Shuffle}, who first proposed to introduce randomization in ranking, several works have employed randomization as a means to collect unbiased implicit feedback from user behavior data~\cite{JoachimsSS18UnbiasLTR, abs-cs-0605037ClickthroughLogs, RadlinskiJ07Active, WangBMN16Selection, OosterhuisR20PolicyAware} and for training unbiased ranking models on biased user feedback~\citep{joachims2017unbiased}.
Stochastic ranking policies can also be employed to improve the diversity of retrieved results~\citep{radlinski2008learning} and---as is more relevant to our current work---to ensure fairer exposure of information content~\citep{yadav2019policy,singh:fair-pg-rank,diaz2020evaluating}.
Motivated by these use cases, several recent works~\citep{bruch2020stochastic,yadav2019policy,singh:fair-pg-rank,diaz2020evaluating,OosterhuisR20PolicyAware} have explored optimization of stochastic ranking policies.
In \cref{sec:optimization}, we will demonstrate how stochastic ranking policies for recommendation can be optimized towards JME-fairness.

%% file: prelims.tex
\section{Preliminaries}
\label{sec:prelim}

Before we formally define the different JME-fairness metrics, we go over some of the fundamental concepts in this section that have already been developed in existing literature that we build on in subsequent sections.

\subsection{Exposure and user browsing models}
\label{sec:prelim-user-model}

Simplified models of how users inspect and interact with retrieved results provide a useful tool for metric development~\citep{yilmaz2010expected, chuklin2013click} and for estimating relevance from historical logs of user behavior data~\citep{chuklin2015click}.
These user browsing models provide a mechanism to estimate the probability of exposure of an item $d$ in a retrieved ranked list of items $\sigma$.
For example, the user browsing model behind the rank-biased precision (RBP) metric~\citep{rbp} assumes that the probability of the exposure event $\epsilon$ for $d$ depends only on its rank $\rho_{d,\sigma}$ in $\sigma$ and decreases exponentially as we go down the ranked list.
\begin{align}
    p_{\scaleto{RBP}{3pt}}(\epsilon|d,\sigma) &= \gamma^{(\rho_{d,\sigma} - 1)}
	\label{eqn:rbp-user-model},
\end{align}
where the $\gamma$ is the patience factor and controls how deep in the ranking the user is likely browse.

In this work, we employ the RBP user browsing model although other models can also be employed in its place.

\subsection{Stochastic ranking and expected exposure}
\label{sec:prelim-expected-exposure}

\citet{diaz2020evaluating} define a stochastic ranking policy $\pi$ as a probability distribution over all permutations of items in the collection.
In the recommendation scenario, let $\mathcal{D}$ be the collection of all items and $\mathcal{U}$ the set of all users of the retrieval system.
Now, given a stochastic ranking policy $\pi_u$ conditioned on user $u \in \mathcal{U}$, we can compute the expected value of the probability that an item $d \in \mathcal{D}$ is exposed to the user as follow:
\begin{align}
    p(\epsilon|d, \pi_u) &= \mathbb{E}_{\sigma \sim \pi_u}\left[p(\epsilon|d,\sigma)\right].
	\label{eqn:expected-exposure}
\end{align}

Assuming the RBP user browsing model, we can further compute $p(d|\sigma)$ based on \cref{eqn:rbp-user-model}.
Furthermore, for notational convenience, let $\mathsf{E} \in \mathbb{R}^{|\mathcal{U}| \times |\mathcal{D}|}$ be the expected exposure matrix, such that $\mathsf{E}_{ij} = p(\epsilon|\mathcal{D}_j, \pi_{\mathcal{U}_i})$.
In the rest of this paper, we refer to the expected exposure $\mathsf{E}$ corresponding to a stochastic ranking policy $\pi$ as determined by a retrieval system as \emph{system exposure}.
Similarly, \emph{target exposure} is defined as the expected exposure $\mathsf{E}^*$ corresponding to an ideal stochastic ranking policy $\pi^*$, whose behavior may be dictated by some desirable principle, such as the \emph{equal expected exposure principle}~\citep{diaz2020evaluating}.
The deviation of $\mathsf{E}$ from $\mathsf{E}^*$ gives us a quantitative measure of the suboptimality of the retrieval system under consideration.
Finally, we define \emph{random exposure} as the expected exposure $\mathsf{E}^\sim$ corresponding to a stochastic ranking policy $\pi^\sim$ which is defined by a uniformly random distribution over all permutations of items in the collection.

%% file: jme.tex
\section{Joint multisided exposure (JME) fairness}
\label{sec:jme}

\begin{figure*}[t]
    \centering
    \begin{subfigure}[t]{0.23\linewidth}
        \includegraphics[width=\textwidth]{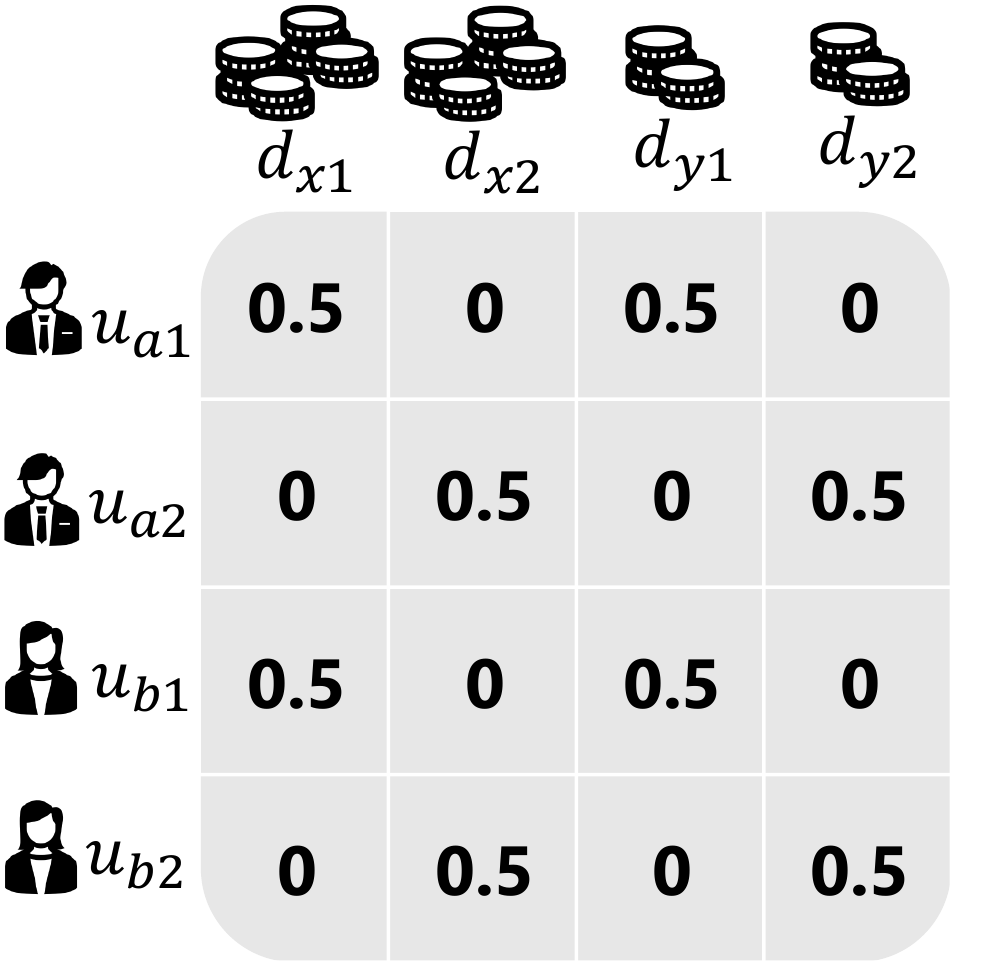}
        \caption{II-F=$0.0625$, IG-F=$0$, GI-F=$0$, GG-F=$0$, AI-F=$0$, AG-F=$0$}
        \label{fig:toy-example-1}
    \end{subfigure}
    \hspace{5em}
    \begin{subfigure}[t]{0.23\linewidth}
        \includegraphics[width=\textwidth]{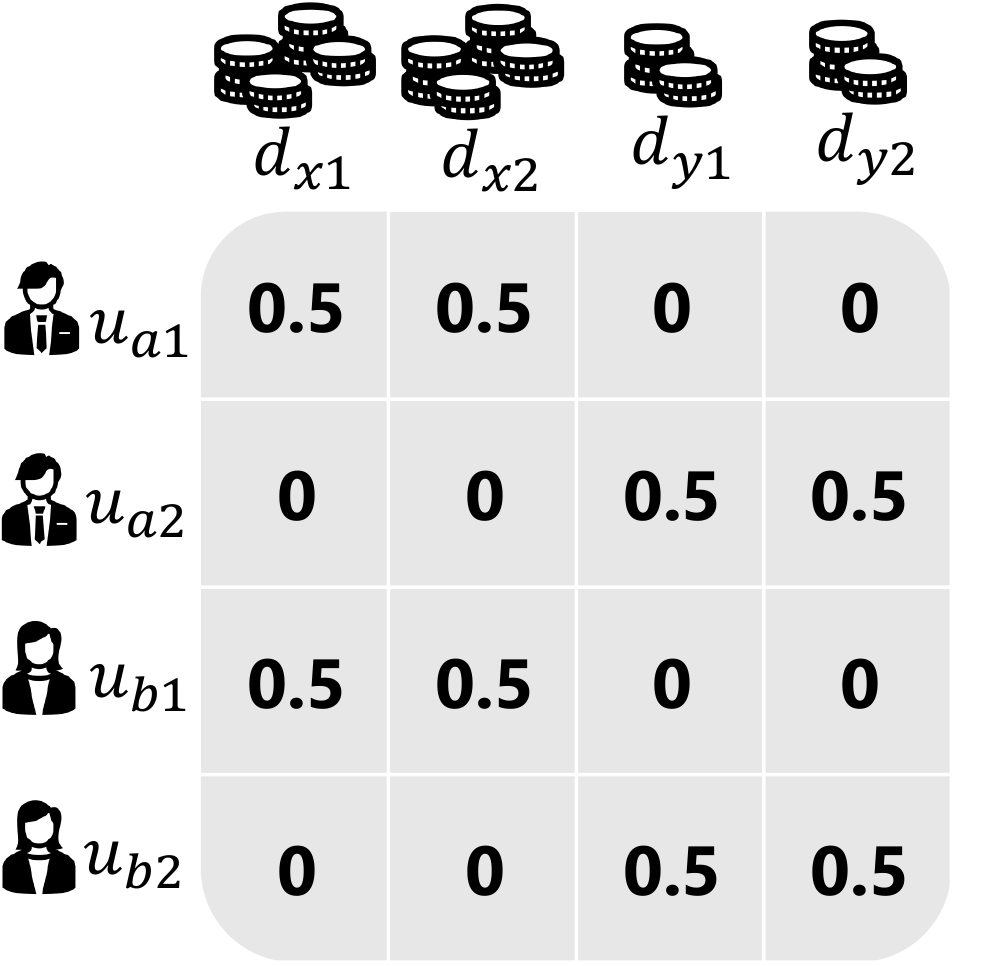}
        \caption{II-F=$0.0625$, IG-F=$0.0625$, GI-F=$0$, GG-F=$0$, AI-F=$0$, AG-F=$0$}
        \label{fig:toy-example-2}
    \end{subfigure}
    \hspace{5em}
    \begin{subfigure}[t]{0.23\linewidth}
        \includegraphics[width=\textwidth]{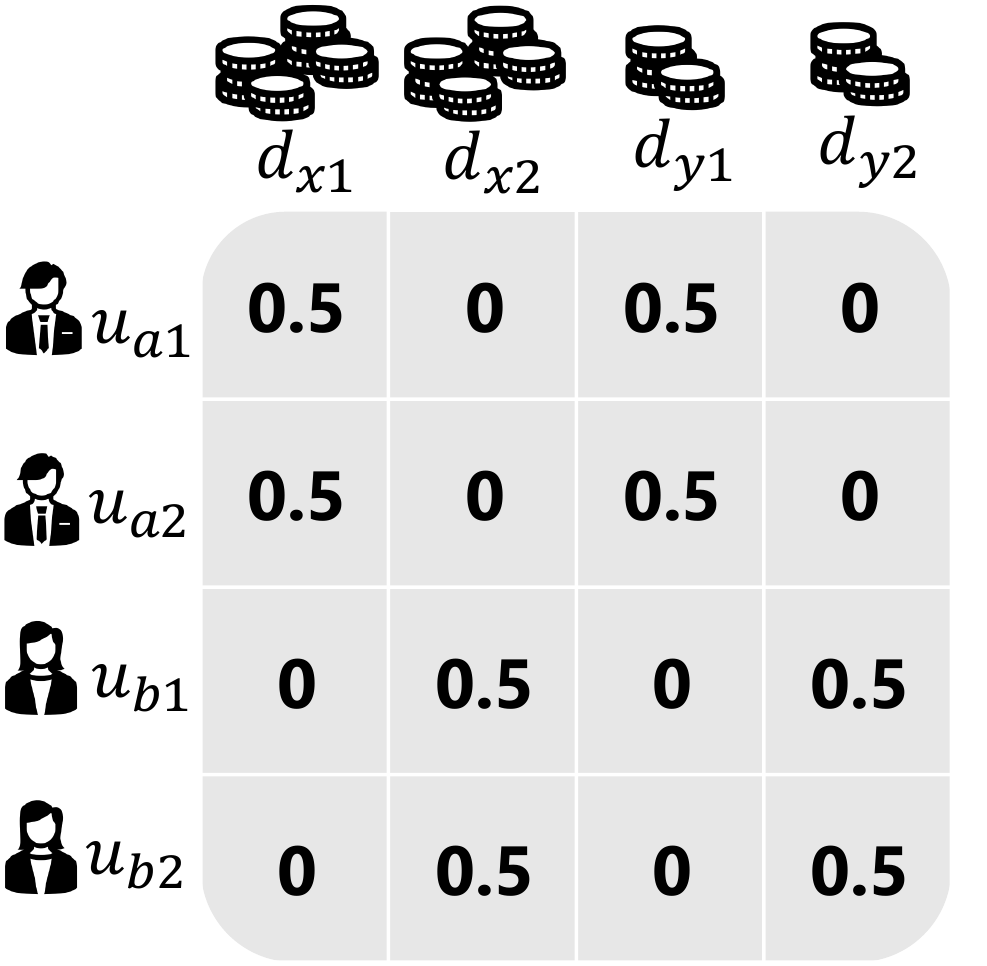}
        \caption{II-F=$0.0625$, IG-F=$0.0$, GI-F=$0.0625$, GG-F=$0$, AI-F=$0$, AG-F=$0$}
        \label{fig:toy-example-3}
    \end{subfigure}
    
    \begin{subfigure}[t]{0.23\linewidth}
        \includegraphics[width=\textwidth]{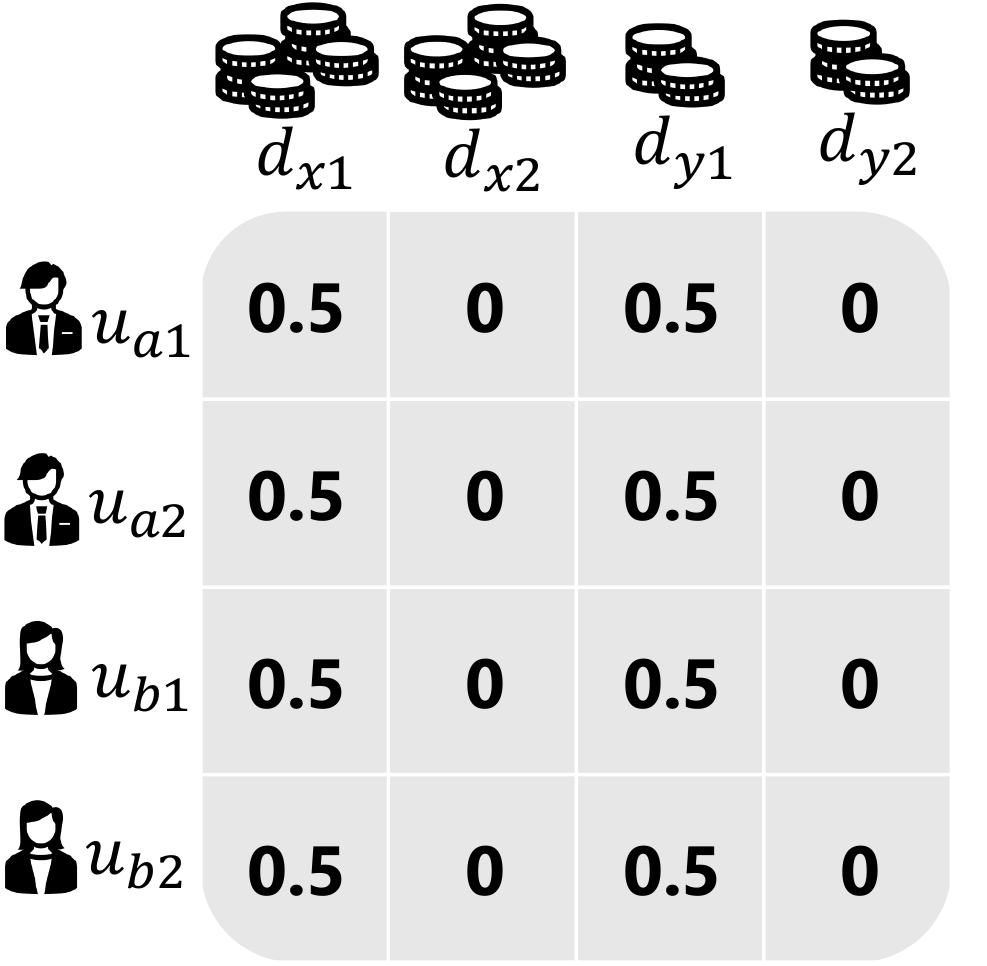}
        \caption{II-F=$0.0625$, IG-F=$0.0$, GI-F=$0.0625$, GG-F=$0$, AI-F=$0.0625$, AG-F=$0$}
        \label{fig:toy-example-4}
    \end{subfigure}
    \hspace{5em}
    \begin{subfigure}[t]{0.23\linewidth}
        \includegraphics[width=\textwidth]{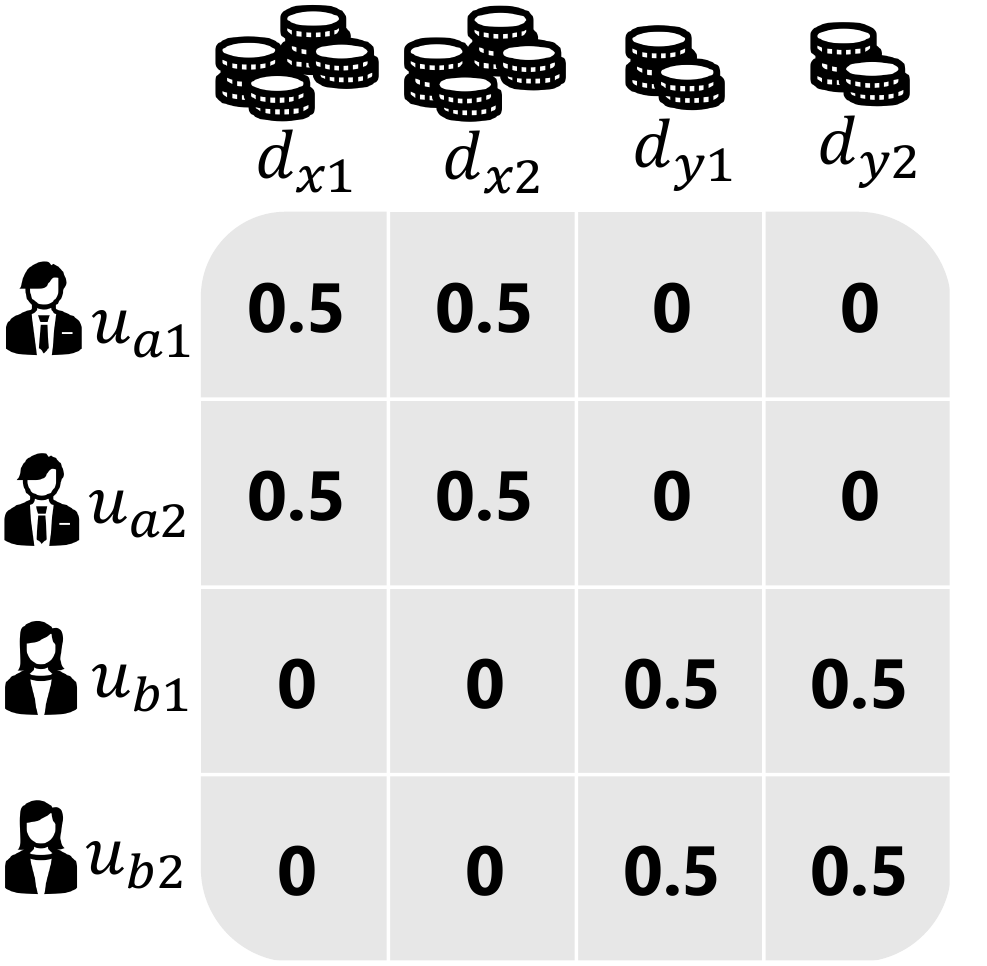}
        \caption{II-F=$0.0625$, IG-F=$0.0625$, GI-F=$0.0625$, GG-F=$0.0625$, AI-F=$0$, AG-F=$0$}
        \label{fig:toy-example-5}
    \end{subfigure}
    \hspace{5em}
    \begin{subfigure}[t]{0.23\linewidth}
        \includegraphics[width=\textwidth]{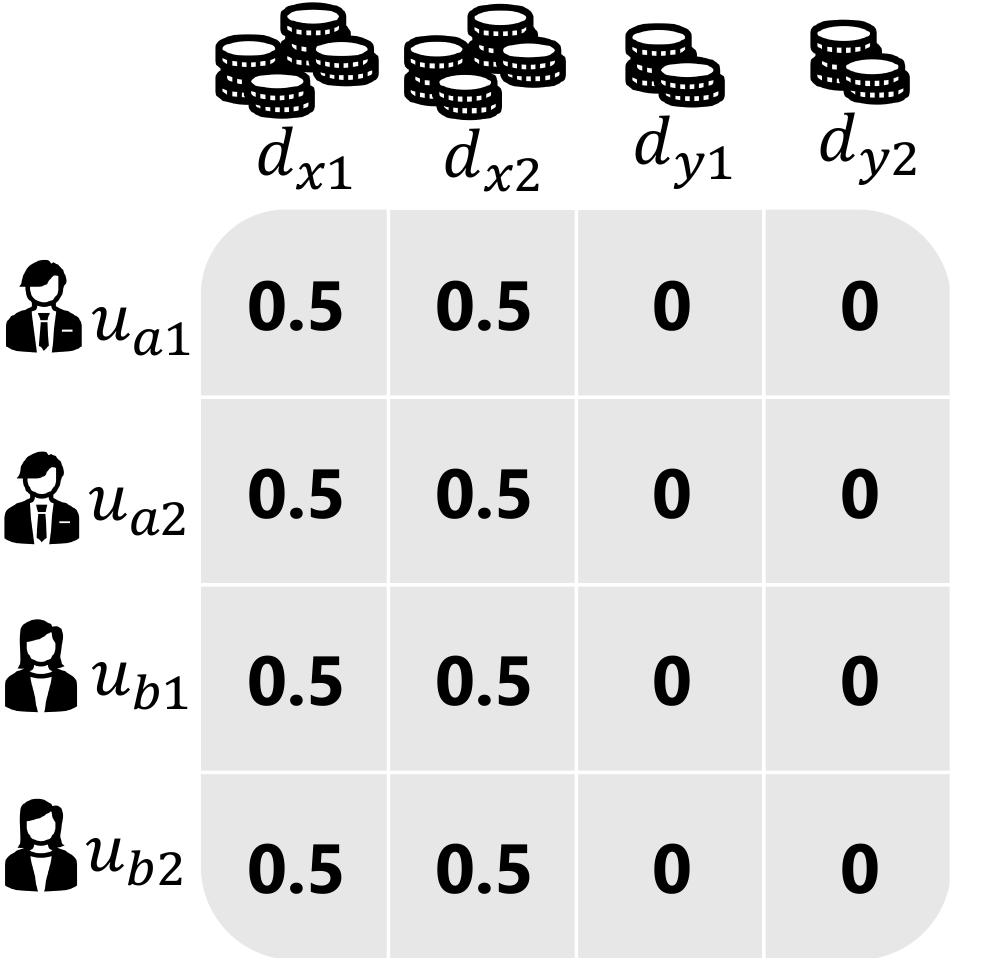}
        \caption{II-F=$0.0625$, IG-F=$0.0625$, GI-F=$0.0625$, GG-F=$0.0625$, AI-F=$0.0625$, AG-F=$0.0625$}
        \label{fig:toy-example-6}
    \end{subfigure}
    \caption{A toy example of a job recommendation system demonstrating the different fairness concerns.
    We present six systems that have comparable II-F metric values but differ in their fairness as measured along the other five JME-fairness dimensions.
    The values in the tables represent the probability of exposure of an individual job to an individual candidate.}
    \label{fig:toy-example}
\end{figure*}

Having introduced the expected exposure framework in the previous section, we now formalize six different JME-fairness metrics.
Each of these metrics addresses a distinct aspect of fairness, covering the disparity in how individual items and groups of items are exposed to individual users, groups of users, and all users.


\subsection{Metric defintions}
\label{sec:jme-definition}

\paragraph{Individual-user-to-individual-item fairness (II-F).}
The II-F metric, previously proposed by \citet{diaz2020evaluating}, measures the disparity between the system and target exposure at the level of individual users and individual items.
Using similar notations as before, we have:

\begin{small}
\begin{align}
\textbf{II-F} &=\frac{1}{|\mathcal{D}|}\frac{1}{|\mathcal{U}|}\sum_{d\in \mathcal{D}}\sum_{u\in \mathcal{U}}\left(p(\epsilon|d, u)-p^*(\epsilon|d, u)\right)^2 \label{eqn:metric-iif1} \\
&= \frac{1}{|\mathcal{D}|}\frac{1}{|\mathcal{U}|}\sum_{j=1}^{|\mathcal{D}|}\sum_{i=1}^{|\mathcal{U}|}(\mathsf{E}_{ij}-\mathsf{E}^*_{ij})^2. \label{eqn:metric-iif2}
\end{align}
\end{small}

The key motivation of this work lies in the observation that the deviation between the system and target exposure may not be distributed uniformly across different user sub-populations and item groups.
As we will demonstrate in \cref{sec:jme-relationship}, the II-F metric cannot distinguish the scenario where the system-to-target exposure deviations systemically and disproportionately impact specific user and/or item groups from the scenario where these deviations are comparable across groups.

\paragraph{Individual-user-to-group-of-items fairness (IG-F).}
We first introduce group attributes on the item-side and present the IG-F metric which is concerned with whether groups of items are over or under exposed to individual users.
We achieve this by making couple of minor modifications to \cref{eqn:metric-iif1}:
\begin{enumerate*}[label=(\roman*)]
    \item replacing $p(\epsilon|d, u)$ and $p^*(\epsilon|d, u)$ with $p(\epsilon|D, u)$ and $p^*(\epsilon|D, u)$, respectively, where $D \in \mathcal{G}_d$ denotes a group of items and $\mathcal{G}_d$ is the set of all item groups, and
    \item averaging the deviations across groups of items instead of individual items.
\end{enumerate*}

\begin{small}
\begin{align}
\textbf{IG-F} &= \frac{1}{|\mathcal{G}_d|}\frac{1}{|\mathcal{U}|}\sum_{D \in \mathcal{G}_d}\sum_{u \in \mathcal{U}}\left(p(\epsilon|D, u)-p^*(\epsilon|D, u)\right)^2 \label{eqn:metric-igf1} \\
&= \frac{1}{|\mathcal{G}_d|}\frac{1}{|\mathcal{U}|}\sum_{D \in \mathcal{G}_d}\sum_{i=1}^{|\mathcal{U}|}\left(\sum_{j=1}^{|D|}p(D_j|D)(\mathsf{E}_{ij}-\mathsf{E}^*_{ij})\right)^2. \label{eqn:metric-igf2}
\end{align}
\end{small}

Here, $p(D_j|D)$ can be defined as a uniform probability distribution over all items in a group, or when appropriate a popularity weighted distribution over items can also be employed.

In previous work, \citet[\S4.3]{diaz2020evaluating} have also considered group attributes on the producer-side, and their proposed formulation is in fact identical to \cref{eqn:metric-igf2}.
However, they provide limited motivation for their specific formulation and as a minor contribution \cref{eqn:metric-igf1} gives us a probabilistic interpretation of this measure.

\paragraph{Group-of-users-to-individual-item fairness (GI-F).}
Next, we introduce group attributes on the user-side which gives us the GI-F metric that measures the over or under exposure of individual items to groups of users.
Similar to the way we define the IG-F metric, the GI-F metric can be defined as follows, where $U \in \mathcal{G}_u$ denote a group of users and $\mathcal{G}_u$ the set of all user groups:

\begin{small}
\begin{align}
\textbf{GI-F} &=\frac{1}{|\mathcal{D}|}\frac{1}{|\mathcal{G}_u|}\sum_{d\in \mathcal{D}}\sum_{U \in \mathcal{G}_u}\left(p(\epsilon|d, U)-p^*(\epsilon|d, U)\right)^2 \label{eqn:metric-gif1} \\
&= \frac{1}{|\mathcal{D}|}\frac{1}{|\mathcal{G}_u|}\sum_{j=1}^{|\mathcal{D}|}\sum_{U\in \mathcal{G}_u}\left(\sum_{i=1}^{|U|}p(U_i|U)(\mathsf{E}_{ij}-\mathsf{E}^*_{ij})\right)^2. \label{eqn:metric-gif2}
\end{align}
\end{small}

Consequently, $p(U_i|U)$ can be defined as a uniform probability distribution over all users in a group, or could be proportional to their usage of the recommender system.

\paragraph{Group-of-users-to-group-of-items fairness (GG-F).}
Having introduced group attributes for users and items separately, we now change our focus to exposure disparities that emerge when we look at group attributes for both the users and items jointly.
Using similar notations as before, we can write:

\begin{small}
\begin{align}
\textbf{GG-F} &= \frac{1}{|\mathcal{G}_d|}\frac{1}{|\mathcal{G}_u|}\sum_{D \in \mathcal{G}_d}\sum_{U \in \mathcal{G}_u}\left(p(\epsilon|D,U)-p^*(\epsilon|D,U)\right)^2 \label{eqn:metric-ggf1} \\
&= \frac{1}{|\mathcal{G}_d|}\frac{1}{|\mathcal{G}_u|}\sum_{D \in \mathcal{G}_d}\sum_{U \in \mathcal{G}_u}\left(\sum_{j=1}^{|D|}\sum_{i=1}^{|U|} p(D_j|D) p(U_i|U)(\mathsf{E}_{ij}-\mathsf{E}^*_{ij})\right)^2. \label{eqn:metric-ggf2}
\end{align}
\end{small}

Of all six fairness metrics defined in this section, the GG-F metric is particularly interesting as all the other metrics can be thought of specific instances of GG-F.
For example, if we define the group attributes for users in GG-F such that each group contains only one user and every user belongs to only one group then we recover the IG-F metric.
A similar trivial definition of groups on the item-side gives us the GI-F metric.
Consequently, if this trivial definition of groups is applied to both the users and items, we get the II-F metric.
Another trivial, but conceptually interesting, definition of the user group may involve a single group to which all users belong.
Under this setting, depending on group definition on the item-side, we can recover the AI-F and AG-F metrics that we describe next.

\paragraph{All-users-to-individual-item fairness (AI-F).}
A recommender system may systemically under or over expose an item to all users.
To quantify this kind of systemic disparities we define the AI-F metric which computes the mean deviation between overall system exposure $p(\epsilon|d)$ and target exposure $p^*(\epsilon|d)$ for items:

\begin{small}
\begin{align}
\textbf{AI-F} &= \frac{1}{|\mathcal{D}|}\sum_{d \in \mathcal{D}}\left(p(\epsilon|d)-p^*(\epsilon|d)\right)^2 \label{eqn:metric-aif1} \\
&= \sum_{j=1}^{|\mathcal{D}|}\left(\sum_{i=1}^{|\mathcal{U}|} p(\mathcal{U}_i)(\mathsf{E}_{ij}-\mathsf{E}^*_{ij})\right)^2. \label{eqn:metric-aif2}
\end{align}
\end{small}

As earlier, $p(\mathcal{U}_i)$ can either be uniform or weighted by usage.

\paragraph{All-users-to-group-of-items fairness (AG-F).}
Finally, the AG-F metric is concerned with systemic under or over exposure of groups of items to all users and is defined as follows:

\begin{small}
\begin{align}
\textbf{AG-F} &= \frac{1}{|\mathcal{G}_d|}\sum_{D \in \mathcal{G}_d}\left(p(\epsilon|D)-p^*(\epsilon|D)\right)^2 \label{eqn:metric-agf1} \\
&= \frac{1}{|\mathcal{G}_d|}\sum_{D \in \mathcal{G}_d}\left(\sum_{j=1}^{|D|}\sum_{i=1}^{|\mathcal{U}|} p(D_j|D)  p(\mathcal{U}_i)(\mathsf{E}_{ij}-\mathsf{E}^*_{ij})\right)^2. \label{eqn:metric-agf2}
\end{align}
\end{small}

\bigskip\noindent
We have formally defined six JME-fairness metrics---II-F, IG-F, GI-F, GG-F, AI-F, and AG-F.
Readers should note that for all six metrics, a \textit{lower value} is more desirable and corresponds to a more fair recommendation.
Next, we discuss the relationship and distinction between these different fairness notions to help the reader develop a more intuitive understanding of our proposed framework of metrics for exposure fairness.

\begin{table*}
\footnotesize
\centering
\caption{Decomposing each of the six JME-fairness metrics into their disparity (*-D) and relevance (*-R) components. }
\vspace{1mm}
\begin{tabular}{l l}
\toprule
{\normalsize \textbf{Disparity}} & {\normalsize \textbf{Relevance}} \\
\midrule
$\begin{aligned}
    \textbf{II-D} &= \frac{1}{|\mathcal{D}|}\frac{1}{|\mathcal{U}|}\sum_{j=1}^{|\mathcal{D}|}\sum_{i=1}^{|\mathcal{U}|}{\mathsf{E}^\delta}_{ij}^2
\end{aligned}$ &
$\begin{aligned}
    \textbf{II-R} &= \frac{1}{|\mathcal{D}|}\frac{1}{|\mathcal{U}|}\sum_{j=1}^{|\mathcal{D}|}\sum_{i=1}^{|\mathcal{U}|}2{\mathsf{E}^\delta}_{ij}{\mathsf{E}^\Delta}_{ij}
\end{aligned}$ \\
$\begin{aligned}
    \textbf{IG-D} &= \frac{1}{|\mathcal{G}_d|}\frac{1}{|\mathcal{U}|}\sum_{D \in \mathcal{G}_d}\sum_{i=1}^{|\mathcal{U}|}\left(\sum_{j=1}^{|D|}p(D_j|D){\mathsf{E}^\delta}_{ij}\right)^2
\end{aligned}$ &
$\begin{aligned}
    \textbf{IG-R} &= \frac{1}{|\mathcal{G}_d|}\frac{1}{|\mathcal{U}|}\sum_{D \in \mathcal{G}_d}\sum_{i=1}^{|\mathcal{U}|}\left(\sum_{j=1}^{|D|}2p(D_j|D){\mathsf{E}^\delta}_{ij}{\mathsf{E}^\Delta}_{ij}\right)^2
\end{aligned}$ \\
$\begin{aligned}
    \textbf{GI-D} &= \frac{1}{|\mathcal{D}|}\frac{1}{|\mathcal{G}_u|}\sum_{j=1}^{|\mathcal{D}|}\sum_{U\in \mathcal{G}_u}\left(\sum_{i=1}^{|U|}p(U_i|U){\mathsf{E}^\delta}_{ij}\right)^2
\end{aligned}$ &
$\begin{aligned}
    \textbf{GI-R} &= \frac{1}{|\mathcal{D}|}\frac{1}{|\mathcal{G}_u|}\sum_{j=1}^{|\mathcal{D}|}\sum_{U\in \mathcal{G}_u}\left(\sum_{i=1}^{|U|}2p(U_i|U){\mathsf{E}^\delta}_{ij}{\mathsf{E}^\Delta}_{ij}\right)^2
\end{aligned}$ \\
$\begin{aligned}
    \textbf{GG-D} &= \frac{1}{|\mathcal{G}_d|}\frac{1}{|\mathcal{G}_u|}\sum_{D \in \mathcal{G}_d}\sum_{U \in \mathcal{G}_u}\left(\sum_{j=1}^{|D|}\sum_{i=1}^{|U|} p(D_j|D)  p(U_i|U){\mathsf{E}^\delta}_{ij}\right)^2
\end{aligned}$ &
$\begin{aligned}
    \textbf{GG-R} &= \frac{1}{|\mathcal{G}_d|}\frac{1}{|\mathcal{G}_u|}\sum_{D \in \mathcal{G}_d}\sum_{U \in \mathcal{G}_u}\left(\sum_{j=1}^{|D|}\sum_{i=1}^{|U|} 2  p(D_j|D)  p(U_i|U){\mathsf{E}^\delta}_{ij}{\mathsf{E}^\Delta}_{ij}\right)^2
\end{aligned}$ \\
$\begin{aligned}
    \textbf{AI-D} &= \sum_{j=1}^{|\mathcal{D}|}\left(\sum_{i=1}^{|\mathcal{U}|} p(\mathcal{U}_i){\mathsf{E}^\delta}_{ij}\right)^2
\end{aligned}$ &
$\begin{aligned}
    \textbf{AI-R} &= \sum_{j=1}^{|\mathcal{D}|}\left(\sum_{i=1}^{|\mathcal{U}|} 2 p(\mathcal{U}_i){\mathsf{E}^\delta}_{ij}{\mathsf{E}^\Delta}_{ij}\right)^2
\end{aligned}$ \\
$\begin{aligned}
    \textbf{AG-D} &= \frac{1}{|\mathcal{G}_d|}\sum_{D \in \mathcal{G}_d}\left(\sum_{j=1}^{|D|}\sum_{i=1}^{|\mathcal{U}|} p(D_j|D)  p(\mathcal{U}_i){\mathsf{E}^\delta}_{ij}\right)^2
\end{aligned}$ &
$\begin{aligned}
    \textbf{AG-R} &= \frac{1}{|\mathcal{G}_d|}\sum_{D \in \mathcal{G}_d}\left(\sum_{j=1}^{|D|}\sum_{i=1}^{|\mathcal{U}|} 2 p(D_j|D)  p(\mathcal{U}_i){\mathsf{E}^\delta}_{ij}{\mathsf{E}^\Delta}_{ij}\right)^2
\end{aligned}$ \\
\bottomrule
\end{tabular}
\label{tab:tab:dr-components}
\end{table*}

\subsection{Relationship between metrics}
\label{sec:jme-relationship}
If we look closely at \cref{eqn:metric-iif2,eqn:metric-igf2,eqn:metric-gif2,eqn:metric-ggf2,eqn:metric-aif2,eqn:metric-agf2}, we notice that all six JME-fairness metrics consider the system-to-target exposure deviations $(\mathsf{E}_{ij}-\mathsf{E}^*_{ij})$ for individual items to individual users.
However, they differ in how they aggregate these differences across users and items leading to interesting relationships and distinctions between these metrics.

Firstly, it is easy to demonstrate that a system that is fair with respect to II-F will also be fair along the other five JME-fairness dimensions.
This is because II-F can be zero if and only if $(\mathsf{E}_{ij}-\mathsf{E}^*_{ij}) = 0, \forall i \in [1, ..., |\mathcal{U}|], j \in [1, ..., |\mathcal{D}|]$, and when that is true then IG-F, GI-F, GG-F, AI-F, and AG-F will also be zero.
However, the reverse does not hold---\ie, a system that may be considered fair based on one or more metrics from the set IG-F, GI-F, GG-F, AI-F, and AG-F may not necessarily be II-fair.
Similarly, it also holds that if a system is either IG-fair or GI-fair then it must also be GG-fair.
To demonstrate this, we have to consider that IG-F is zero if and only if $(\sum_{j=1}^{|D|}p(D_j|D)(\mathsf{E}_{ij}-\mathsf{E}^*_{ij})) = 0, \forall i \in [1, ..., |\mathcal{U}|]$ which implies that $(\sum_{i=1}^{|U|}p(U_i|U)\sum_{j=1}^{|D|} p(D_j|D)  (\mathsf{E}_{ij}-\mathsf{E}^*_{ij})) = 0, \forall U_i \in U, U \in \mathcal{G}_u$.
Similarly, we can also prove the relationship between GI-F and GG-F.
As before, a system that is GG-fair, however, may not necessarily be IG-fair or GI-fair.
Furthermore, we can also show that a system will necessarily be AI-fair if it is GI-fair, and similarly both GG-fairness and AI-fairness independently imply AG-fairness.
We do not include the full proof for each of these relationships due space constraints but leave them as an useful exercise for the reader.

Next, we consider the question that if II-fairness implies fairness along the other five dimensions then why it is necessary to develop these other fairness notions at all.
The answer lies in the fact that two systems that are comparable according to the II-F metric, may demonstrate very different levels of unfairness along the other dimensions.
To motivate this more intuitively, let us consider the toy example of a job recommender system in \cref{fig:toy-example}.
In this example, there are four candidates ($u_{a1}$, $u_{a2}$, $u_{b1}$, $u_{b2}$) and four jobs ($d_{x1}$, $d_{x2}$, $d_{y1}$, $d_{y2}$), and all four jobs are relevant to each of the four candidates.
The candidates belong to two groups $a$ ($u_{a1}$ and $u_{a2}$) and $b$ ($u_{b1}$ and $u_{b2}$)---\eg, based on gender---and similarly the jobs belong to two groups $x$ ($d_{x1}$ and $d_{x2}$) and $y$ ($d_{y1}$ and $d_{y2}$), say based on whether they pay high or low salaries.
Furthermore, we assume that the recommender system displays only one result at a time and our simple user model assumes that the user always inspects the displayed result---\ie, the probability of exposure is 1 for the displayed item and 0 for all other items for a given impression.
In this setting, an ideal recommender would expose each of the four jobs to each candidate with a probability of $0.25$.

The six fictitious recommender systems shown in~\cref{fig:toy-example}, demonstrate comparable disparity according to the II-F metric but different levels of unfairness along the other dimensions.
For example \cref{fig:toy-example-1,fig:toy-example-3,fig:toy-example-4} correspond to systems that are IG-fair because every candidate is exposed to at least one high-paying and one low-paying job, whereas in \cref{fig:toy-example-2,fig:toy-example-5,fig:toy-example-6} every candidate is exposed to either only high-paying or only low-paying jobs raising IG-unfairness concerns.
Similarly, we can observe that \cref{fig:toy-example-1,fig:toy-example-2} are GI-fair because each group of candidates are exposed to all four jobs, while consequently the other four systems demonstrate GI-unfairness.
Among these six examples, only \cref{fig:toy-example-5,fig:toy-example-6} demonstrate GG-unfairness.
In \cref{fig:toy-example-5}, men are recommended high-paying jobs while women are recommended low-paying jobs, and in \cref{fig:toy-example-6} all candidates are exposed only to high-paying jobs leading to no candidates being referred to low-paying jobs.
Finally, \cref{fig:toy-example-4,fig:toy-example-6} demonstrate AI-unfairness because they include individual jobs that are under or over exposed to all candidates, and specifically in the latter instance we also observe AG-unfairness as the low-paying jobs are under-exposed to all candidates as previously noted.
It is precisely these differences that motivate us to develop these six JME-fairness metrics which we believe measures different systemic unfairness in recommendation outcomes, and therefore should be considered in complementary combination for more holistic fairness analysis.

\begin{figure*}[t]
    \centering
    \includegraphics[width=\linewidth]{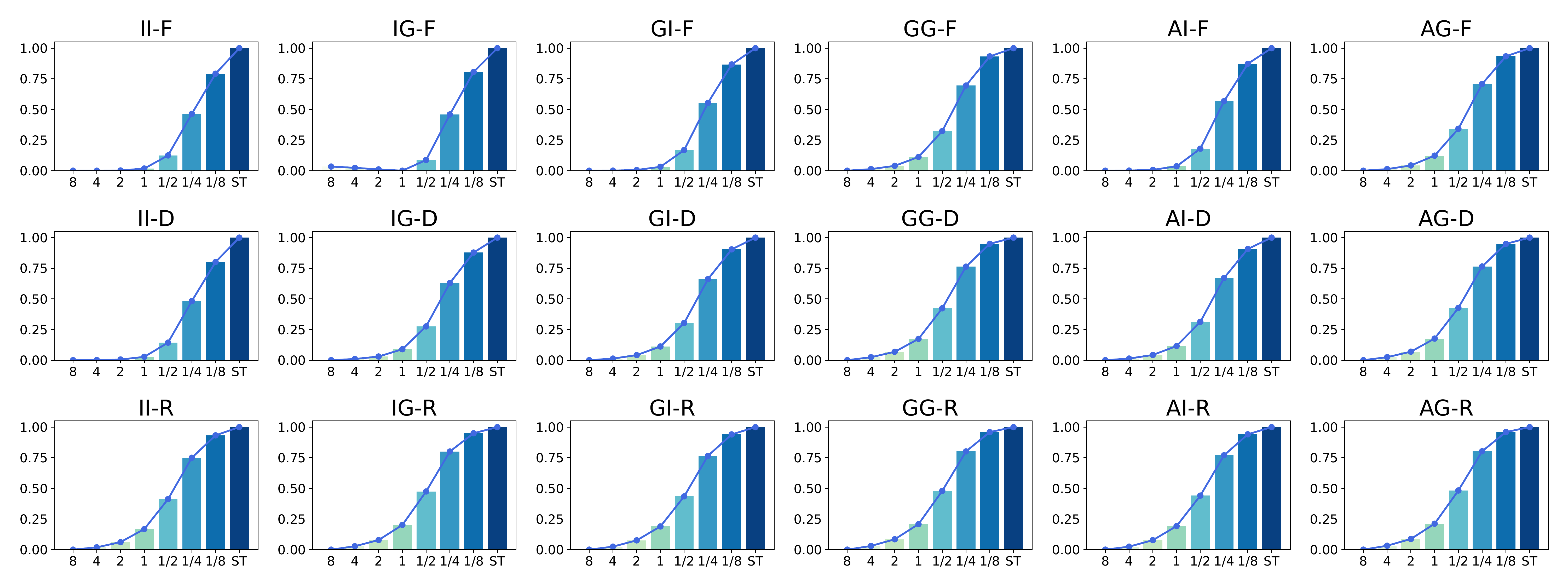}
    \caption{Behavior of JME-fairness metrics for a stochastic ranking policy---generated by randomizing the BPRMF model using Plackett-Luce---on the \textit{MovieLens1M} dataset. The first row shows the impact of different stochasticity on the overall fairness metrics, while the second and third row show the impact on their corresponding disparity and relevance components. The x-axis shows the values of $\beta$, where a larger value indicates more randomization.}
    \label{fig:result-randomization-impact}
\end{figure*}

\subsection{Disparity and relevance}
\label{sec:jme-components}

\citet{diaz2020evaluating} make an interesting observation that as an artifact of using squared error to compute the deviation between the system and target exposure. The II-F metric can be decomposed into a \emph{disparity} component (II-D), a \emph{relevance} component (II-R), and a remaining component that is a system-independent constant (II-C).

\begin{small}
\begin{align}
    \textbf{II-F} =& \frac{1}{|\mathcal{D}|}\frac{1}{|\mathcal{U}|}\sum_{j=1}^{|\mathcal{D}|}\sum_{i=1}^{|\mathcal{U}|}(\mathsf{E}_{ij}-\mathsf{E}^*_{ij})^2 & \nonumber \\
    =& \frac{1}{|\mathcal{D}|}\frac{1}{|\mathcal{U}|}\sum_{j=1}^{|\mathcal{D}|}\sum_{i=1}^{|\mathcal{U}|}\mathsf{E}_{ij}^2 &\Big\} \text{II-D} \nonumber \\
    &- \frac{1}{|\mathcal{D}|}\frac{1}{|\mathcal{U}|}\sum_{j=1}^{|\mathcal{D}|}\sum_{i=1}^{|\mathcal{U}|}2\mathsf{E}_{ij}\mathsf{E}^*_{ij} &\Big\} \text{II-R} \nonumber \\
    &+ \frac{1}{|\mathcal{D}|}\frac{1}{|\mathcal{U}|}\sum_{j=1}^{|\mathcal{D}|}\sum_{i=1}^{|\mathcal{U}|}{\mathsf{E}^*}_{ij}^2 &\Big\} \text{II-C} \nonumber
\end{align}
\end{small}

This decomposition is useful as it represents an inherent trade-off between the disparity and relevance as stochasticity is introduced into the model.
Increasing randomization in the model has the desirable property of reducing disparity (II-D) but also the undesirable effect of reducing relevance (II-R).
In this framing, relevance is maximized when a static ranking model is employed and disparity is minimized when the ranking model is fully stochastic.

A similar decomposition into disparity and relevance components would also be useful for the new metrics proposed in this paper.
When considering group attributes, say on the item-side, it should be self-evident that a fully stochastic ranking model would recommend items not necessarily with uniform probability over the groups, but in proportion to the group sizes in the collection.
In \cref{sec:prelim-expected-exposure}, we defined random exposure $\mathsf{E}^\sim$ as the expected exposure matrix corresponding to such a fully stochastic model.
It makes intuitive sense to define disparity not in terms of the flatness of the system exposure distribution but as a deviation of the system exposure from random exposure, especially when the distribution under consideration is over groups of items rather than individual items.
Therefore, we propose a slight modification to the decomposition of the original II-F metric, which is then consistent with how we decompose the other five JME-fairness metrics into their respective components.
To do so, we first rewrite \cref{eqn:metric-iif2} as follows:

\begin{small}
\begin{align}
    \textbf{II-F} &= \frac{1}{|\mathcal{D}|}\frac{1}{|\mathcal{U}|}\sum_{j=1}^{|\mathcal{D}|}\sum_{i=1}^{|\mathcal{U}|}(\mathsf{E}_{ij}-\mathsf{E}^*_{ij})^2 \nonumber \\
    &= \frac{1}{|\mathcal{D}|}\frac{1}{|\mathcal{U}|}\sum_{j=1}^{|\mathcal{D}|}\sum_{i=1}^{|\mathcal{U}|}\big((\mathsf{E}_{ij}-\mathsf{E}^\sim_{ij}) - (\mathsf{E}^*_{ij}-\mathsf{E}^\sim_{ij})\big)^2. \label{eqn:metric-iif3}
\end{align}
\end{small}

For notational brevity, let $\mathsf{E}^\delta_{ij} = \mathsf{E}_{ij}-\mathsf{E}^\sim_{ij}$ and $\mathsf{E}^\Delta_{ij} = \mathsf{E}^*_{ij}-\mathsf{E}^\sim_{ij}$.
Based on \cref{eqn:metric-iif3}, we now redefine II-D and II-R as:

\begin{small}
\begin{align}
    \textbf{II-D} &= \frac{1}{|\mathcal{D}|}\frac{1}{|\mathcal{U}|}\sum_{j=1}^{|\mathcal{D}|}\sum_{i=1}^{|\mathcal{U}|}{\mathsf{E}^\delta}_{ij}^2 \label{eqn:metric-iid} \\
    \textbf{II-R} &= \frac{1}{|\mathcal{D}|}\frac{1}{|\mathcal{U}|}\sum_{j=1}^{|\mathcal{D}|}\sum_{i=1}^{|\mathcal{U}|}2{\mathsf{E}^\delta}_{ij}{\mathsf{E}^\Delta}_{ij}. \label{eqn:metric-iir}
\end{align}
\end{small}

Consequently, the new definition of II-D would produce the same system ordering as the original definition by \citet{diaz2020evaluating}.
Due to space constraints, we do not include the full proof here.
Next, employing a similar strategy we can decompose the other five JME-fairness metrics into their disparity and relevance components as shown in \cref{tab:tab:dr-components}.
In \cref{sec:analysis-results}, we will analyse how stochasticy trades-off between the disparity and relevance components of all six JME-fairness metrics.

%% file: analysis.tex
\section{Metric analysis}
In this section, we analyze the proposed six JME-fairness metrics with respect to the trade-offs involved between their respective disparity and relevance components (intra-metric analysis) and the relationship between the different JME-fairness metrics themselves (cross-metric analysis).
Specifically, we will study the following research questions:
\begin{itemize}
    \item \textbf{RQ1}: What is the impact of stochasticity on each of the six JME-fairness metrics, and their corresponding disparity and relevance components?
    \item \textbf{RQ2}: How do the different JME-fairness metrics, and their respective components, correlate with each other and what does that tell us about the relationships between these different fairness dimensions?
\end{itemize}
\label{sec:analysis}

\begin{figure*}[t]
    \centering
    \includegraphics[width=\linewidth]{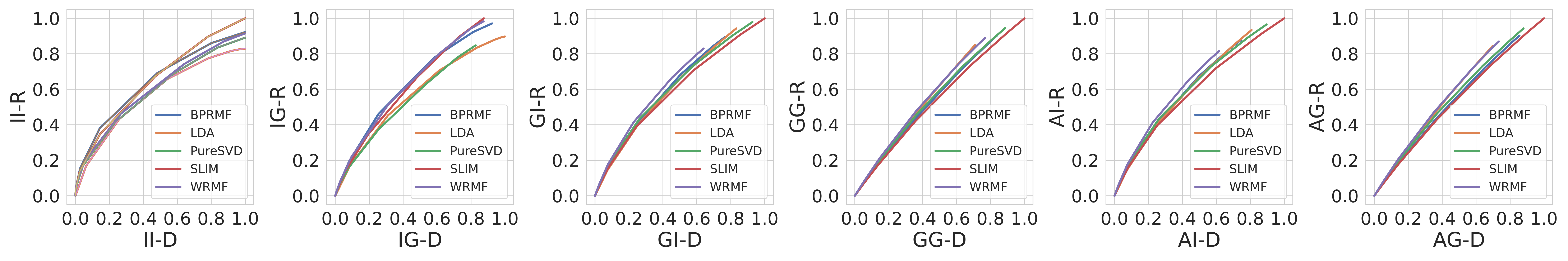}
    \caption{Curves for disparity-relevance trade-off across six different fairness dimensions. We introduce different level of stochaticity on the top of the static rankings from six different recommendation models.}
    \label{fig:result-dr-tradeoff}
\end{figure*}

\subsection{Method}
\label{sec:analysis-method}
As aforementioned, we are interested in studying stochastic ranking policies with respect to their ability to distribute exposure across (individual or groups of) items.
Towards that goal, we first introduce a method for generating stochastic ranking policies based on a deterministic ranking model.
Given a deterministic ranker, and corresponding estimated relevance scores for items with respect to user $u$, we utilize the Plackett-Luce (PL) model~\citep{Luce59PL1959, plackett1975analysis} to sample multiple rankings.
The PL model depends on Luce’s axiom of choice which states that the odds of choosing an item over another do not depend on the set of items from which the choice is made~\cite{Luce59PL1959, LUCE1977215PL1977}.
Specifically, the PL model constructs a ranking by iteratively sampling items without replacement from the collection with probability distribution $p(d|u)$ defined as below:
\begin{align}
    p(d|u)=\frac{\exp(\bm{Y}_{d,u}/\beta)}{\sum_{d' \in \mathcal{D}}\exp(\bm{Y}_{d',u}/\beta)},
\label{eq:rand_PL}
\end{align}
where $\bm{Y}_{d,u}$ is the relevance score estimated by the deterministic ranker for item $d$ with respect to user $u$.
The parameter $\beta$ is the softmax temperature.
A larger $\beta$ corresponds to more stochasticity in the ranking.
For example, when $\beta=8$, the probability distribution over all permutations is almost uniform and the stochastic policy approaches a fully random ranking model.
As a corollary, when $\beta$ decreases the stochastic policy converges to the deterministic ranking policy, which is a ranking of items sorted by their estimated relevance score $\bm{Y}_{d,u}$ in descending order for each user $u$.

We generate stochastic ranking policies by applying this post-processing step, with different values of $\beta$, over a set of trained recommendation models that are publicly available~\footnote{We obtain the set of runs from: \href{https://github.com/dvalcarce/evalMetrics}{https://github.com/dvalcarce/evalMetrics}.} for the \textit{MovieLens1M} dataset.
There are $6,040$ users and $3,706$ items in the \textit{MovieLens1M} dataset.
For the group information, we use the gender and age attributes on the user side, and the genre attributes on the item side.
We rerank the top $100$ items and sample $100$ rankings for each user during evaluation.
We employ the RBP user browsing model and set the patience factor $\gamma=0.8$.
We select different values for $\beta$ in the range of $1/8$ to $8$ for introducing different degree of stochasticity in our ranking, and compare with the deterministic ranking policy.

\subsection{Results}
\label{sec:analysis-results}

\paragraph{RQ1: Impact of Stochasticity.}

To investigate the impact of stochasticity, we first visualize how different values of $\beta$ influences the different JME-fairness metrics and their corresponding components in \cref{fig:result-randomization-impact}.
Our analysis is based on a stochastic ranking policy that uses the BPRMF model as the underlying deterministic ranker, although we have confirmed that we obtain similar results when considering other ranking models.
For consistency, we normalize each of the metric values between 0 and 1 using min-max normalization.
Our first key observation is that as $\beta$ increases, the values on all metrics tend to decrease.
This is expected given that a larger $\beta$ corresponds to a more random ranking policy, where the original static relevance estimates have a smaller influence, which consequently results in both lower disparity and relevance.
As a corollary, the static ranking policy derived directly from the deterministic BPRMF model has the highest relevance and disparity for all six fairness metrics.

For each of the JME-fairness metrics, these results imply the existence of a disparity-relevance trade-off as the stochastic ranking policy cannot achieve both low disparity and high relevance simultaneously.
To visualize this more clearly, we present the disparity-relevance trade-off curves for all six JME-fairness metrics in \cref{fig:result-dr-tradeoff}.
We study five different models in this experiment: BPRMF~\cite{RendleFGS09bprmf}, LDA~\cite{BleiNJ03LDA}, PureSVD~\cite{HuKV08PureSVD}, SLIM~\cite{NingK11SLIM}, and WRMF~\cite{TakacsPNT09WRMF}, employing the same randomization strategy on top for all models.
As before, we normalize the disparity and relevance metric values in the range of 0 to 1.
A point closer to the lower left represents a ranking policy with full stochasticity, while the right-most point for each curve corresponds to the respective static ranking policy---\ie, without any randomization.

For a given metric, different models achieve different levels of disparity-relevance trade-off, and which model achieves the best trade-off may differ across the different JME-fairness metrics.
For a given metric, this can be best quantified by computing the area under the disparity-relevance curve for each model, as shown in \cref{tab:auc-model-comparison}.
For example, WRMF model achieves the best trade-off for GI-F, GG-F, AI-F and AG-F, but is worse for II-F compared to SLIM and BPRMF.

\begin{table}[t]
\centering
\caption{The AUC of the disparity-relevance trade-off curves for different models across the six fairness dimensions. The highest value in each column is bolded, while the second highest is underlined.}
\vspace{1mm}
\resizebox{0.48\textwidth}{!}{
\begin{tabular}{lrrrrrr}
\toprule
Model & II & IG & GI & GG & AI & AG\\
\midrule
BPRMF & \underline{0.6331} & \textbf{0.4774} & 0.2904 & 0.2953 & 0.2712 & 0.2814\\
LDA& 0.5664 & 0.4088 & 0.2837 & \underline{0.3164} & 0.2687 & \underline{0.3115}\\
PureSVD & 0.5830 & 0.4102 & \underline{0.2921} & 0.3030 & \underline{0.2755} & 0.2942\\
SLIM & \textbf{0.6408} & 0.4654 & 0.2776 & 0.2851 & 0.2605 & 0.2752\\
WRMF & 0.5996 & \underline{0.4769} & \textbf{0.3135} & \textbf{0.3186} & \textbf{0.2957} & \textbf{0.3139}\\
\bottomrule
\end{tabular}
}
\label{tab:auc-model-comparison}
\vspace{-3mm}
\end{table}

\begin{figure}
    \centering
    \includegraphics[width=\linewidth]{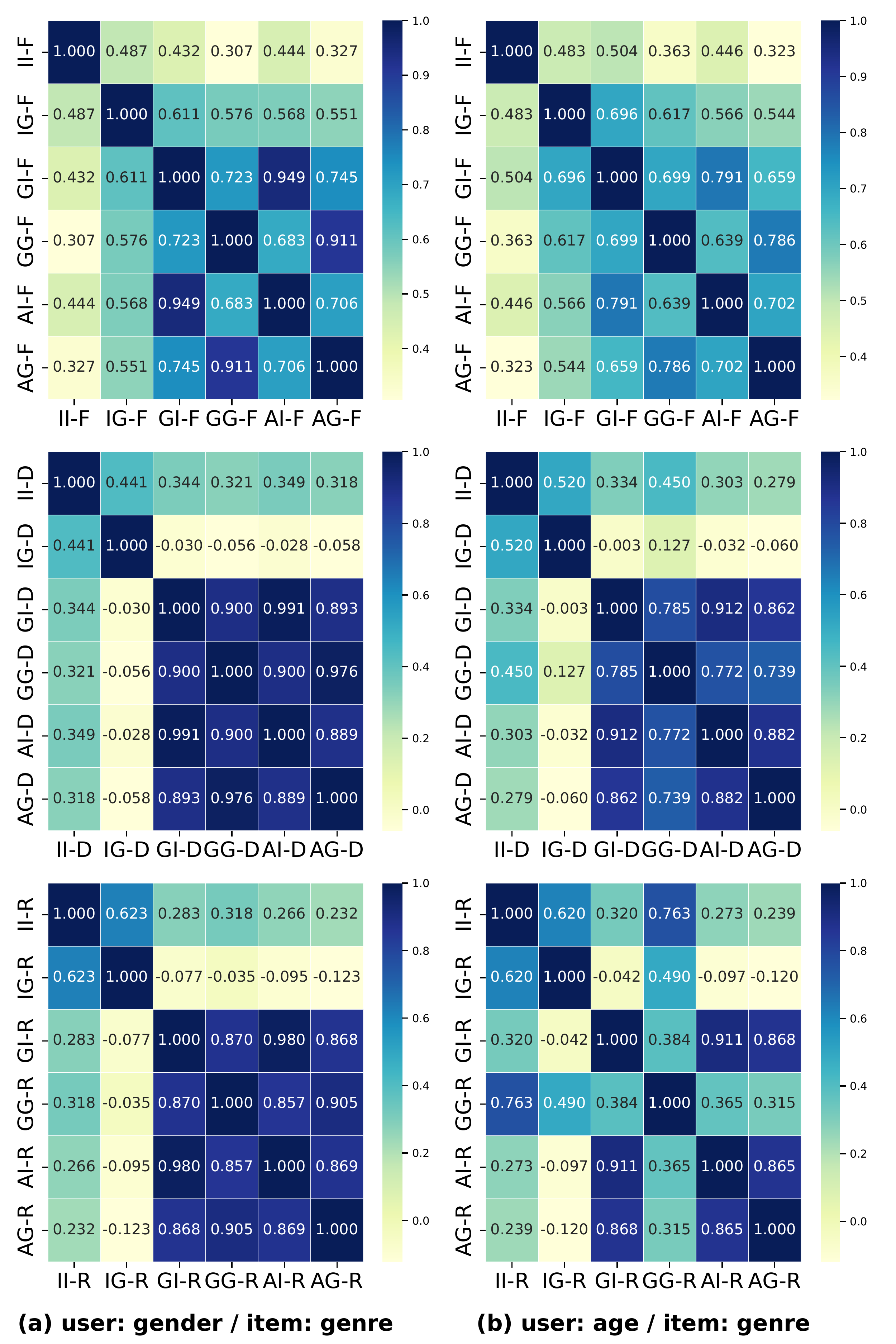}
    \vspace{0.05mm}
    \caption{The Kendall rank correlation between different metrics and their two components across six different fairness dimensions.}
    \label{fig:result-metric-correlation}
\vspace{-2mm}
\end{figure}

\paragraph{RQ2: Metric Correlation.} 
The intra-metric analysis in the previous subsection demonstrates how all six JME-fairness metrics trades-off between their corresponding disparity and relevance components.
Next, we shift our focus to cross-metric analysis to understand how a recommender system optimized for one of the JME-fairness metrics may fare on the other measures.
To study this, we evaluate $15$ different recommendation models---with $7$ different levels of stochasticity in each case---against each of the six JME-fairness metrics.
For each metric, this gives us $15 \times 7$ values each corresponding to a recommender system instance---\ie, a combination of model and stochasticity level.
Now for every pair of JME-fairness metrics, we compute the Kendall rank correlation~\cite{kendall1948rank} to quantify the agreement between the two metrics with respect to the ordering of the recommender system instances, as presented in \cref{fig:result-metric-correlation}.
We perform the analysis twice using gender and age as demographic attributes for the user-side, respectively.

We observe that the correlation between the II-F metric and the other five JME-fairness measures is typically low, and the same holds when we compare the corresponding disparity and relevance components across these measures.
This provides evidence to our claim that each of these metrics quantifies different notion of unfairness, and that a system instance that may perform well on one fairness dimension may be suboptimal for another.
A more nuanced interpretation of these correlation matrices requires us to recall that in \cref{sec:jme-definition} we argued that the the other five JME-fairness metrics can be considered as a specific instance of the GG-F metric based on the group definitions.
For example, on the user dimension, if our group definition involves a very small number of groups then the GG-F metric should intuitively display high correlation with the AG-F metric.
On the other hand, if we consider many small groups of users in context of GG-F, then we would expect the correlation between GG-F and IG-F to be higher.
In \cref{fig:result-metric-correlation}, the group definition based on both gender and age correspond to the scenario where we have a small number of groups which may explain why we see strong correlation between GG-F and AG-F---and between age and gender we see that the correlation is higher for gender because it involves relatively fewer groups compared to age.
Effectively, GG-F captures a notion of unfairness that lies on spectrum between what IG-F and AG-F measures.
The correlation matrix therefore is consistent with the behavior we would expect by comparing \cref{eqn:metric-iif2,eqn:metric-igf2,eqn:metric-gif2,eqn:metric-ggf2,eqn:metric-aif2,eqn:metric-agf2}.
\vspace{-1mm}

%% file: optimization.tex
\section{Optimization for JME-fairness}
\label{sec:optimization}
In the previous section, we demonstrated that a recommender system that performs well on one JME-fairness metric may under-perform on another.
This highlights a potential opportunity to optimize multiple JME-fairness metrics simultaneously.
Ideally, we expect that optimizing for multiple JME-fairness measures allows us to explicitly trade-off between these various fairness concerns.


\subsection{Algorithm}
\label{sec:optimization-algo}
To optimize towards a JME-fairness objective, we adopt the approach proposed by \citet{diaz2020evaluating}.
Let $\bm{Y}\in\mathbb{R}^{|\mathcal{U}|\times|\mathcal{D}|}$ be the estimated relevance between users and items as predicted by a model $f_\theta: \mathcal{U} \times \mathcal{D} \to \mathbb{R}$, parameterized by $\Theta$.
A stochastic ranking policy can now be defined that sampling items without replacement from the following probability distribution as given by the Plackett-Luce model:

\begin{align}
    p(\mathcal{D}_j|\mathcal{U}_i)=\frac{\exp(\bm{Y}_{ij})}{\sum_{k}\exp(\bm{Y}_{ik})}.
\label{eq:PL}
\end{align}

Sampling from the above distribution is a non-differentiable step that creates a roadblock for gradient-based optimization.
However, following the same strategy as \citet{diaz2020evaluating}, we can reparameterize the probability distribution by adding independently-drawn noise $\zeta$ sampled from the Gumbel distribution to $\bm{Y}_{ij}$, and sorting the items by their ``noisy'' probability distribution $\tilde{p}(\mathcal{D}_j|\mathcal{U}_i)$: 

\begin{align}
    \tilde{p}(\mathcal{D}_j|\mathcal{U}_i)=\frac{\exp(\bm{Y}_{ij}+\zeta_j)}{\sum_{k}\exp(\bm{Y}_{ik}+\zeta_k)}.
\label{eq:PL_gumbel}
\end{align} 

The sorting step itself is also non-differentiable, but we can compute the smooth rank~\cite{WuCZZ09SmoothDCG,qin2010general} for each item in the ranking as follows:
\begin{equation}
    \rho_{\mathcal{D}_j, \pi_{
    \mathcal{U}_i}} = \sum_{k \in [1..|\mathcal{D}|], k\neq j}\left(1+\exp\left(\frac{ \tilde{p}(\mathcal{D}_j|\mathcal{U}_i)- \tilde{p}(\mathcal{D}_k|\mathcal{U}_i)}{\tau}\right)\right)^{-1} \,,
\label{eq:smooth_rank}
\end{equation}
where the temperature $\tau$ is a hyperparameter that controls the smoothness of the approximated ranks.
We have now computed the rank position of an item $\mathcal{D}_j$ in the ranking with respect to user $\mathcal{U}_i$ in a differentiable way.
Next, we can compute the system exposure using a user browsing model like RBP as $\mathsf{E}_{ij}=(1-\gamma)\cdot\gamma^{\rho_{\mathcal{D}_j, \pi_{\mathcal{U}_i}}-1}$.
To derive the expected exposure, we average the system exposure over $100$ different sampled rankings.
Finally, having estimated the system exposure $\mathsf{E}$ we can now compute its deviation from target exposure $\mathsf{E}^*$ using different JME-fairness metric definitions which are themselves differentiable.

Readers should note that merely optimizing towards a JME-fairness metric, such as GG-F, is not sufficient for a recommender system to learn a relevance function between individual users and items.
Therefore, it is important to combine the II-F objective, which is appropriate for relevance modeling, with the other desired JME-fairness objective(s).
In this work, we combine the GG-F objective linearly with the II-F as follows:
\begin{align}
    \mathcal{L} = \text{II-F}+\alpha\cdot \text{GG-F},
\label{eq:opt_obj}
\end{align}
where $\alpha$ is a scaling factor that trades-off between the two objectives.
Similarly, other combinations of JME-fairness notions can also be developed.

\subsection{Experiment}
\label{sec:optimization-experiment}

\paragraph{Datasets.}
Since our proposed JME-fairness metrics require meaningful attribute information on both the user-side and item-side, we select the \textit{MovieLens100K} and \textit{MovieLens1M} \cite{HarperK16movielens} as our datasets.
Both of these datasets contain user-to-movie relevance information collected from the \textit{MovieLens} website.
These two datasets provide 100 thousand and 1 million user-movie interactions, respectively, with the user metadata (gender, age, occupation group) and movie genres.
For each user, we randomly select 20\% of the rated items as ground truth for testing, The remaining 70\% and 10\% data constitutes the training and validation set. 
~\footnote{The code and datasets are available at \href{https://github.com/haolun-wu/JMEFairness}{https://github.com/haolun-wu/JMEFairness}.} 

\paragraph{Implementation Details.} 
We choose Matrix Factorization (MF) as our base model to conduct our experiments.
MF is widely used in recommendation, due to its simplicity and efficiency.
We optimize the model using the Adam optimizer with the Xavier initialization~\cite{GlorotB10xavier}.
The model parameters $\Theta$ contain the user embeddings and item embeddings, where the embedding size is fixed to 64.
In order to obtain the user-item relevance matrix $\bm{Y}$, we use the inner-product to compute the similarity between each pair of users and items.
The batch size for training is set to be 32.
The learning rate and the regularization hyper-parameter are selected from $ \{1e^{-2}, 1e^{-3}, 1e^{-4}\} $.
The scaling factor $\alpha$ is selected from \{0, 1, 5, 10, 20 ,50\}.
The temperature $\tau$ in \cref{eq:smooth_rank} and the patience parameter $\gamma$ for the RBP user browsing model are set to be 0.1 and 0.8, respectively.

\begin{figure}
    \centering
    \includegraphics[width=\linewidth]{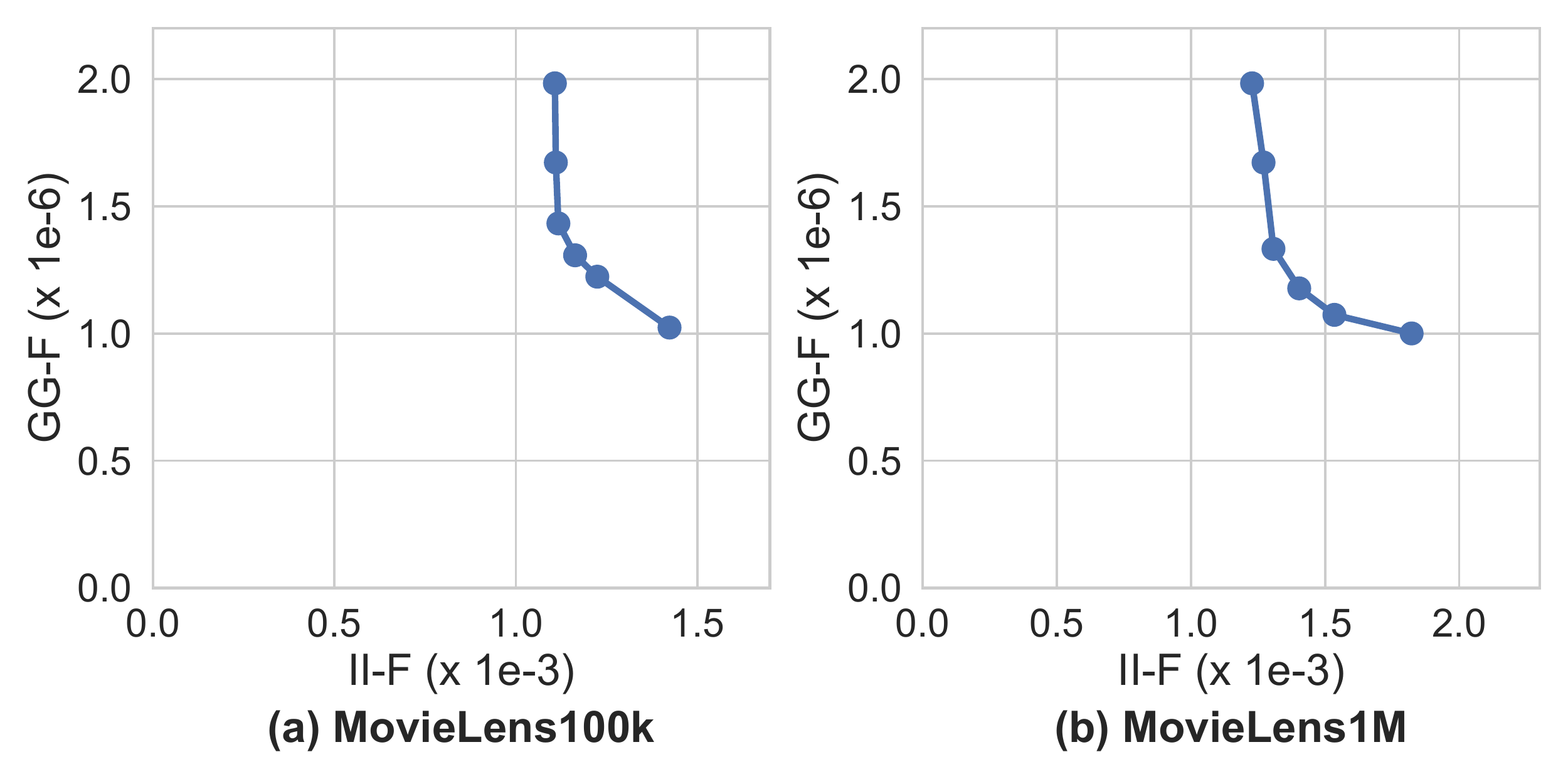}
    \caption{The trade-off between II-F and GG-F when directly optimizing the recommender model towards a combination of both objectives.}
    \label{fig:optimization-tradeoff}
\end{figure}

\subsection{Results}
\label{sec:optimization-results}

\Cref{fig:optimization-tradeoff} shows the trade-off between II-F and GG-F that we achieve by varying $\alpha$.
On both datasets, we initially see a sharp fall in GG-F, which is desirable, with only a small increase in II-F, which depending on the application may be a fair trade-off against the GG-F improvements.
For example, if we compare training the model with $\alpha=0$ (\ie, training on II-F only) and $\alpha=1$, we see a statistically significant improvement (drop) in GG-F according to the student's t-test ($p<0.01$), while the corresponding degradation (increase) in II-F is not statistically significant under the same test, on both datasets.
Furthermore, the degradation in NDCG between $\alpha=0$ and $\alpha=1$ is also small on both datasets, as shown in \cref{tab:rec-quality}, and the difference is not statistically significant under the student's t-test ($p<0.01$).
These empirical results portray the opportunity to optimize recommender systems along multiple exposure fairness dimensions, while maintaining high recommendation quality.



%% file: conclusion.tex
\section{Discussion and Conclusion}
\label{sec:conclusion}
In IR, many measures of retrieval quality focus primarily on the position of relevant items in static ranked lists.
By explicitly measuring deviations in system and target exposure, expected exposure metrics, in contrast, provide a useful framework to consider other dimensions of ranking evaluations, such as diversity and fairness, in addition to relevance.
While, this framework have previously been employed in the context of individual users to individual and groups of items exposure fairness, we argue in this work that joint consideration of group attributes on both user-side and item-side allows us to study other forms of systematic unfairness of social and moral import in recommendation.
In this setting, the choice of group attributes on both sides is an important consideration and must be informed by historical and social contexts as well as critical scholarship in the area of socioeconomic justice.
For example, in the context of job recommendation, such considerations may include an understanding of historical and ongoing pay and other forms of workplace discrimination based on gender and race~\citep{daugherty2021history, williams2021wage}.
While, the analysis presented in this paper has considered group attributes for users (and correspondingly for items) along single dimensions, it is noteworthy that our formalization does not make any such assumptions.
In fact, in the job recommendation scenario, it is meaningful to consider, say, both gender and race attributes of applicants jointly with grouping of jobs by salary and demographic attributes of corresponding business-owners in the GG-F metric formulation.
Incorporating multiple group dimensions also raises additional considerations, such as of intersectional fairness~\citep{crenshaw2017intersectionality, collins2020intersectionality}.
We have not yet analysed the appropriateness and sufficiency of our proposed framework for quantifying unfairness that may result from membership in multiple historically marginalized demographics but we believe this is an exciting area for future studies.

\begin{table}[t]
\centering
\caption{NDCG@50 for the recommender system corresponding to different values of $\alpha$.}
\vspace{1mm}
\begin{tabular}{lrrrrrr}
\toprule
Dataset & $\alpha=0$ & $\alpha=1$ & $\alpha=5$ & $\alpha=10$ & $\alpha=20$ & $\alpha=50$\\
\midrule
\textit{ML100K} & 0.3703 & 0.3692 & 0.3684 & 0.3680 & 0.3674 & 0.3662 \\
\textit{ML1M} & 0.2741 & 0.2736 & 0.2725 & 0.2712 & 0.2693 & 0.2684\\
\bottomrule
\end{tabular}
\label{tab:rec-quality}
\vspace{-5mm}
\end{table}

A historical perspective on fairness and justice is also critical to ensuring appropriate operationalization of our proposed framework.
For example, readers should note that computing target exposure assumes the availability of \emph{true relevance} labels.
However, all real-world recommendation datasets likely suffer from historical biases~\citep{mehrabi2021survey} and therefore computing target exposure based on these \emph{observed relevance} labels is also likely to be unfair to different demographics.
In some cases, this gap between true and observed relevance may be mitigated by modifying how target exposure is computed---\eg, for GG-F it may be reasonable to employ $\mathsf{E}^\sim$ instead of $\mathsf{E}^*$ as the ideal target exposure distribution over item groups.
The application of our proposed framework of metrics necessitates such careful deliberation to prevent the undesirable case of bias-in-bias-out.

Our work also has implications for the design and deployment of stochastic ranking models.
Considering multiple fairness metrics, such as those proposed in this work, requires trading-off different forms of unfairness---\eg, II-F \vs GG-F.
This may have implications for model optimization as well as the calibration of stochasticity in the model.
While both our current work and previous literature on expected exposure metrics have considered stochastic models that randomize the output rankings by sampling noise from model-independent distributions, future development of stochastic ranking models may also want to consider approaches where the randomization is informed by the model's own epistemic and aleatoric uncertainty~\citep{cohen2021not}. On a concluding note, while our work has developed the framework of JME-fairness in the context of recommendation task, we believe our formulations can easily be extended to other IR tasks, such as search by replacing users with queries, without loss of generality.
